\documentclass[aps,pra,10pt,twocolumn,notitlepage]{revtex4-1}
\usepackage{graphicx}
\usepackage{amssymb}
\usepackage{amsfonts}
\usepackage{amsmath}
\usepackage{amsbsy}
\usepackage{natbib}
\usepackage{comment}
\usepackage[colorlinks=true, urlcolor=blue, citecolor=blue,
linkcolor=blue]{hyperref}
\usepackage{color}
\usepackage[utf8]{inputenc}
\usepackage[T1]{fontenc}
\usepackage{amsthm}
\usepackage{mathtools}
\usepackage{scalerel}

\usepackage{lmodern}
\usepackage{graphicx}
\usepackage{mathtools}
\usepackage{tipa}

\begin{document}

\title{Spatio-temporal delay in photoionization by polarization structured laser fields}

\author{Jonas W\"{a}tzel and Jamal Berakdar}
\affiliation{Institute for Physics, Martin-Luther-University Halle-Wittenberg, 06099 Halle, Germany}

\begin{abstract}
 Focused laser fields with textured polarization  on a sub-wavelength scale allow  extracting information on electronic processes which are not directly accessible by homogeneous fields. Here, we consider photoionization in radially and azimuthally polarized laser fields and show that, due to the local polarizations of the transversal and the longitudinal laser electric field components, the detected photoelectron spectra  depend on the atom position from which the electron originates. The calculation results reported here show the angular dependence of the photoionization amplitude, the quantum phase, and the time delay of the valence shell electrons as the parent atom's position varies across the laser spot. We discuss the possibility of using the photoelectron spectra to identify the position of the ionized atom within the laser spot on the sub-wavelength scale. The proposal and its illustrations underline the potential for the detection of the atomic position based on attosecond
streaking methods.
\end{abstract}

\maketitle

\section{Introduction}
Short laser pulses allow a temporal tracking of the dynamics of  charged particles, down to the attosecond time scale  \cite{krausz2009attosecond}. Spatially structured fields are becoming also increasingly important with a number of existing and prospect applications \cite{rubinsztein2016roadmap}. Laser fields, i.e., propagating electromagnetic waves, can be spatially sculptured both in polarization state and spatial phase affecting the amount of spin angular momentum (SAM) and orbital angular momentum (OAM) densities carried by the field.  When interacting with matter, these fields may lead to new effects, such as unidirectional charge currents \cite{Quinteiro:09, watzel2016centrifugal, sederberg2020Vectorized}, to a first order in the electric field amplitude and without symmetry break, which is interesting for optoelectronics  \cite{solyanik2019spin, PhysRevB.93.045205, KOC2015599, PhysRevB.100.115308, doi:10.10631.5027667, ji2020photocurrent}. For molecules \cite{PhysRevLett.89.143601, PhysRevA.71.055401, PhysRevLett.96.243001}, structured laser pulses are expected to yield new information, particularly on chiral and helical molecular aggregates \cite{PhysRevA.99.023837, Wozniak:19, ayuso2019synthetic} and enable the generation of nano-scale magnetic pulses \cite{watzel2016optical}.\\
A transfer of optical OAM from a phase-structured laser (optical vortex) to electronic states was observed in atoms \cite{de2020photoelectric, schmiegelow2016transfer, afanasev2017circular}.  Electronic processes in atoms driven by polarization-structured laser pulses (also called vector beams) are relatively less studied   \cite{watzel2020electrons, watzel2019magnetoelectric, hernandez2017extreme, watzel2020multipolar, wang2020vectorial}. Another type of field closely related to those discussed here is the propagating optical skyrmions \cite{watzel2020electrons}, which in contrast to vector beams, can imprint OAM on electronic orbitals and have a varying polarization landscape. Here, we show that a superposition of co-propagating SAM-structured vector beams interacting with an atom generates photoelectrons that can carry angular momentum and spatial information with a resolution below the optical diffraction where the atom resides within the laser spot. One of the lasers is a radially polarized, and the other is azimuthally polarized. When tightly focused, the radially polarized beam has a longitudinal linearly polarized field component localized around the optical axis. Away from the optical axis, it is  transversely  radially polarized.  The azimuthally polarized field is purely transverse.
An appropriate combination of both has the following characteristics: i) a  longitudinal field component at the beam center, and ii) circular polarization in the outer beam rim. We infer that the valence electron's dynamics and ejection direction depend markedly on the position from which the photoelectron is launched. As the polarization change occurs on a distance below the wave-length, scanning the photoionization characteristics opens the way to spatial resolution below the diffraction limit. A prominent optical spectroscopy with resolution below optical diffraction limit (down to tens of nanometers resolution) is the stimulated emission depletion (STED) microscopy \cite{klar1999subdiffraction}. There, the resolution is achieved via appropriate intensity modulations or two lasers. A laser field with a donut-shaped spatial intensity profile co-propagates  with a Gaussian laser field. Upon interaction with a sample the fluorescence signals, or the stimulated emission depletion area can be tuned by the intensity ratio of the two laser fields. In contrast, our focus here is on the polarization shaping to gain spatio-temporal spectroscopic information on the electronic structure. Physically, polarization and spatial phase modulations of a laser field are not restricted by the Abbe limit.\\
As measurable quantities, we calculated for an ensemble of Ar atoms the time delay in photoionization \cite{schultze2010delay, ivanov2013time} and its angular dependencies for different atoms in the laser spot, as well as for the ensemble average. A previous study discussed the  spatial resolution via time delay on the sub-wavelength scale by using optical vortices for photoionization \cite{watzel2016discerning}. The idea relied on OAM transfer to the electron. The scheme is an experimental  challenge, however. OAM transfer from vortices to valence electrons is only effective for atoms very close to the optical axis, meaning only a small fraction contributes to the
signal for a moderately dense gas \cite{watzel2020electrons}. Here, we instead exploit the \emph{change} in the polarization state, which has an overall more extensive range but is still on the sub-wavelength limit \cite{watzel2020multipolar}. Furthermore, substantial intensity
resides in the center of the vector beams combination, which substantially increases the efficiency of a possible experiment. \\
Not only the ejection direction and the atomic time delays are affected by the spatially textured polarization. Also the quantum phase of the photoelectron depends on the atom location (origin of the photoelectron) within the laser spot. Therefore, a spatial identification is also possible via attosecond streaking \cite{goulielmakis2004direct} with a space and time-structured laser field. Here, we combine a short circularly polarized XUV pump pulse with a time-delayed spatially inhomogeneous IR probe field. Depending on the position from which the photoelectron is launched, we observe different streaking characteristics at the photoelectron detector. \\
The required  (XUV nad IR) laser fields are feasible. Intense and focussed XUV vector beams can be generated by high harmonic generation \cite{hernandez2017extreme}, for instance via relativistic plasma mirrors \cite{chen2021intense}. For the generation of vector beams in the infrared and visible regime various methods both active and passive do exist, e.g. via tunable q-plates \cite{rumala2013tunable, d2015arbitrary, larocque2016arbitrary}.\\
The rest of the paper is structured as follows: after introducing notation and fields in Sec.II, we discuss in Sec. III the results for the polarization-dependent time delay in Ar and in Sec. IV, we present the spatial resolution based on attosecond streaking, and conclude with Sec.~V.\\
Unless otherwise stated, atomic units (a.u.) are used throughout the text.

\section{Theoretical model}
\subsection{Polarization-varying light mode}
\label{sec:I}
\begin{figure}
\centering
\includegraphics[width=8.5cm]{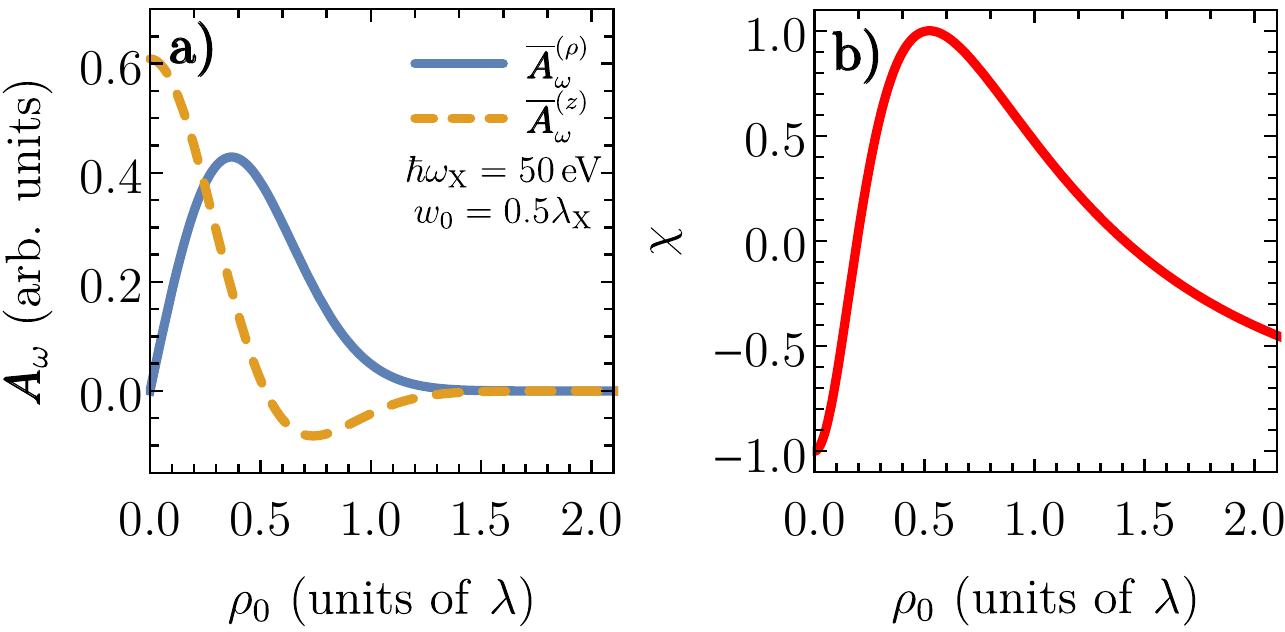}
\caption{Polarization landscape of a linear combination of a radial and phase-shifted azimuthal vector beam. The laser field is focused to the wavelength limit. Panel a) shows the variation  in  magnitudes of the longitudinal and transverse field components as the axial distance $\rho_0$ changes. Panel b) indicates the local polarization as described by the  quantity $\chi$, where $\chi(\rho_0)\rightarrow+1$ ($\chi(\rho_0) \rightarrow-1$ )  corresponds to local circular (linear) polarization.}
\label{Fig1}
\end{figure}

The vector potential of the structured XUV laser field with central frequency $\omega_{\rm X}$  has the Fourier components $\pmb{A}_\omega$, i.e.
$\pmb{A}(\pmb{r},t)=\frac{1}{2\pi}\int{\rm d}\omega\,\pmb{A}_\omega(\pmb{r})e^{-i\omega t},
$
and the spatial coordinate $\pmb{r}$ refers to the optical axis.
For tightly focused beams, $\pmb{A}(\pmb{r},t)$ has  longitudinal and transversal components.
Switching to cylindrical coordinates with the $z$ axis being along the optical axis, the  transverse part of a radially polarized vector beam is given by \cite{zhan2009cylindrical, watzel2019magnetoelectric}
\begin{equation}
\overline{\pmb{A}}_{\omega}^{(\perp)}(\pmb{r})=C_\omega u_q(\rho,z)e^{iqz}\hat{e}_\rho,
\end{equation}
where $C_\omega$ is the Fourier coefficient for the light mode with frequency $\omega$ and $q=\omega_{\rm X}/c$ is the photonic wave vector. The spatial distribution is described by $u_q(r,z)$. For $z\ll z_R$, where $z_R=qw_0^2/2=l/2$ is the Rayleigh length and the parameter $l$ relates to the diffraction length, $u_q(r,z)$ reads \cite{cerjan2011orbital}
\begin{equation}
u_q(\rho,z)=\frac{\rho}{w_0}\exp\left[ -\frac{\rho^2}{w_0^2} + i\frac{qz\rho^2}{l^2} - 4i\frac{z}{l} \right],
\end{equation}
where the beam waist $w_0$ characterizes the extent of the laser spot. It is possible to use the Coulomb gauge, in which case the longitudinal component must obey \cite{quinteiro2019reexamination}
\begin{equation}
\partial_z\overline{\pmb{A}}_{\omega}^{(z)}(\pmb{r},t)= - \pmb{\nabla}_\perp\cdot\overline{\pmb{A}}_{\omega}^{(\perp)}(\pmb{r}).
\end{equation}
For $q>1/l$ (note that $\exp(iqz)$ oscillates much faster than $\exp(iz/l)$) we may approximate
\begin{equation}
\begin{split}
  \overline{A}^{(z)}(\pmb{r},t)&=\frac{i}{q}\pmb{\nabla}_\perp\cdot\overline{\pmb{A}}_\omega^{(\perp)}(\pmb{r}) \\
  &=iC_\omega\frac{A_0}{q}\frac{1}{\rho}\partial_\rho[\rho u_q(\rho,z)]e^{iqz}.
\end{split}
\end{equation}
In the focal plane and around $z=0$, the vector potential fulfilling $\nabla\cdot\overline{\pmb{A}}_\omega(\pmb{r})=0$ reads
\begin{equation}
\overline{\pmb{A}}_{\omega}(\pmb{r})=C_\omega\left[\frac{\rho}{w_0}\hat{e}_\rho + i\frac{2}{qw_0}\left(1-\frac{\rho^2}{w_0^2}\right)\hat{e}_z \right]e^{-\rho^2/w_0^2}e^{iqz}.
\end{equation}
The longitudinal component is dominant around the optical axis. Its relative strength to the transverse component for a large axial distance depends on the focusing condition, as evident from the scaling factor $1/(kw_0)$. The radial field component peaks at $\rho_{\rm max}=w_0/\sqrt{2}$.\\
The transverse vector potential $\widetilde{\pmb{A}}(\pmb{r},t)$ of an azimuthal vector beam fulfills the Coulomb gauge condition and
reads in the vicinity of $z=0$ \cite{zhan2009cylindrical, watzel2019magnetoelectric}:
\begin{equation}
\widetilde{\pmb{A}}_\omega(\pmb{r})=\mathcal{C}_\omega\frac{\rho}{w_0}e^{-\rho^2/w_0^2}e^{iqz}\hat{e}_\varphi.
\end{equation}
For a coherent superposition of the co-propagating radial and azimuthal vector beams, we introduce a (spatially independent) time shift $\Delta t=\pi/(2\omega_{\rm X})$.  For sufficiently long pulses $\mathcal{C}_\omega\cong iC_\omega$ for $\omega\geq0$ and $\mathcal{C}_\omega\cong-iC_\omega$ for $\omega<0$. The combined vector potential in Fourier space is then
\begin{equation}
\begin{split}
\pmb{A}_\omega(\pmb{r})=\,&C_\omega\left[\frac{\rho}{w_0}(\hat{e}_\rho + i\,{\rm sgn}[\omega]\hat{e}_\varphi)\right. \\
& \left. + i\frac{2}{qw_0}\left(1-\frac{\rho^2}{w_0^2}\right)\hat{e}_z \right]e^{-\rho^2/w_0^2}e^{iqz}.
\end{split}
\label{eq:Acomb}
\end{equation}
Due to temporal shift between the azimuthal and radial beam, the total field is \emph{circularly  polarized} in the  transverse  plane, i.e. $\pmb{A}_\omega^{(\perp)}(\pmb{r})\propto(\hat{e}_\rho + i\,{\rm sgn}[\omega]\hat{e}_\varphi)$. This is particularly  relevant for the quantity
\begin{equation}
\chi(\rho_0)=\frac{|\pmb{A}_\omega^{(\perp)}(\rho_0)|^2 - |A_\omega^{(z)}(\rho_0)|^2}{|\pmb{A}_\omega^{(\perp)}(\rho_0)|^2 + |A_\omega^{(z)}(\rho_0)|^2},
\end{equation}
which characterizes the $\emph{local}$ polarization in the laser spot at the axial distance $\rho_0$. For  $\chi(\rho_0)\rightarrow-1$ (cf.~Fig.\,\ref{Fig1}), the field is linearly  polarized (along the $z$-axis) due to  the longitudinal component. $\chi(\rho_0)\rightarrow+1$ indicates  local circular polarization at the axial distance $\rho_0$. \\
Note, radially or azimuthally polarized vector fields do not carry net orbital or spin angular momentum \cite{watzel2019magnetoelectric} and the intensity profiles are cylindrically symmetric. However, the resulting circular polarization in the transversal plane exhibits a dependence on the azimuthal coordinate via a phase $\exp(-i\varphi)$. Therefore, similar fields can be represented by a light wave carrying orbital and oppositely directed spin angular momentum.
\subsection{Photoionization and atomic time delay}
We consider the peak intensity of the constructed vector beam $\pmb{A}(\pmb{r},t)$ corresponding to a Keldysh parameter $\gamma\gg1$, in which case a perturbative treatment of the photoionization process is reasonable \cite{amusia2013atomic}. The temporal envelope of both laser fields is $g(t)=\exp[-t^2/(2\sigma^2)]$ with $\sigma=T_p/(2\sqrt{2\log(2)})$, where $T_p=2n_p\pi/\omega_{\rm X}$ is the pulse length set by the number of  optical cycles $n_p$. The corresponding Fourier coefficients are
\begin{equation}
\begin{split}
C_\omega=\frac{n_p}{2\omega_{\rm X}}\sqrt{\frac{\pi^3}{\log{2}}}
&\left( e^{-\frac{n_p^2\pi^2}{\omega_{\rm X}^2\log(16)}(\omega-\omega_{\rm X})^2} \right. \\
&\left.+ e^{-\frac{n_p^2\pi^2}{\omega_{\rm X}^2\log(16)}(\omega+\omega_{\rm X})^2} \right).
\end{split}
\label{eq:FourierC}
\end{equation}
The description of the constructed beam by Eq.\,\eqref{eq:Acomb} is approximate with a range of validity restricted by  $C_{\omega\rightarrow0}\cong0$. This condition is  fulfilled  for photon energies $\hbar\omega_{\rm X}$ in the XUV regime and $n_p>2$.\\
 Due to the spatial inhomogeneity in both the amplitude and polarization of the laser field, the ionization probability depends  on the atom's position. For an atom located at $\pmb{r}_0=(\rho_0\cos{\varphi_0},\rho_0\sin{\varphi_0},0)^T$,  the axial distance $\rho_0$ to the optical axis is the important parameter. The photoionization amplitude from an initial state $|i\rangle$ upon absorption of one photon from the laser field $\pmb{A}(\pmb{r}_0,t)$ is given by \cite{fedorov1997atomic}
\begin{equation}
\mathcal{A}_{\pmb{k}}(\pmb{r}_0)=-\frac{i}{T}\int_{0}^T{\rm d}t\, \langle\varphi^{(-)}_{\pmb{k}}|\hat{H}_{\rm int}(\pmb{r}_0,t)|\Psi_i\rangle e^{i\omega_{ki}t}.
\label{eq:M1}
\end{equation}
Here, $\langle\pmb{r}|i\rangle=R_{n_i,\ell_i}(r)Y_{\ell_i,m_i}(\Omega_{\pmb{r}}$) is the wave-function of the initial state characterized by the set of quantum numbers $n_i,\ell_i,m_i$ and the orbital energy $E_i$. The angular dependence is characterized by the spherical harmonics $Y_{\ell_i,m_i}(\Omega_{\pmb{r}})$ and the spherical solid angle is defined by $\Omega_{\pmb{r}}=(\vartheta_{\pmb{r}}, \varphi_{\pmb{r}})$. Note,   $\pmb{r}$ is taken with respect to  the atom position at $\pmb{r}_0$. The final state with energy $E_k=k^2/2>0$ is represented by a set of appropriate continuum states  \cite{amusia2013atomic}, i.e.,
\begin{equation}
\langle\pmb{r}|\varphi^{(-)}_{\pmb{k}}\rangle=\sum_{\ell,m}i^\ell e^{-i\delta_\ell(k)}R_{k,\ell}(r)Y_{\ell,m}(\Omega_{\pmb{r}})Y^*_{\ell,m}(\Omega_{\pmb{k}}),
\end{equation}
where $R_{k,\ell}(r)$ are the radial wave functions, $\Omega_{\pmb{k}}=(\vartheta_{\pmb{k}}, \varphi_{\pmb{k}})$ is the spherical solid angle of the momentum $\pmb{k}$, and $\delta_\ell(k)$ are the atom-specific scattering phases representing the scattering characteristics inherent to the atomic potential $V(r)$. In Eq.\,\eqref{eq:M1}, $\omega_{ki}=E_k-E_i$.\\
The atom-lasers interaction is  $\hat{H}_{\rm int}(\pmb{r}_0,t)=\pmb{A}(\pmb{r}_0,t)\cdot\hat{\pmb{p}}$, where $\hat{\pmb{p}}$ is the momentum operator.  Re-expressing
Eq.\,\eqref{eq:M1} by using $\hat{\pmb{p}}=-i[\pmb{r},\hat{H}_0]$ yields \cite{koksal2012charge}
\begin{equation}
\begin{split}
\mathcal{A}_{i\pmb{k}}(\pmb{r}_0)&=\frac{\omega_{ki}}{T}\int_0^T{\rm d}t\, \langle\varphi^{(-)}_{\pmb{k}}|\pmb{A}(\pmb{r}_0,t)\cdot\pmb{r}|\Psi_i\rangle e^{i\omega_{ki}t}\\
&=\omega_{ki}\int{\rm d}\omega\,C_\omega \langle\varphi^{(-)}_{\pmb{k}}|\pmb{A}_\omega(\pmb{r}_0)\cdot\pmb{r}|\Psi_i\rangle \delta_{\omega_{ki},\omega} \\
&=\omega_{ki} C_{\omega_{ki}} \langle\varphi^{(-)}_{\pmb{k}}|\pmb{A}_{\omega_{ki}}(\pmb{r}_0)\cdot\pmb{r}|\Psi_i\rangle,
\end{split}
\label{eq:M2}
\end{equation}
where the Kronecker symbol is written as $\delta_{\omega_{ki},\omega}=\lim_{T\rightarrow\infty}\int_0^T{\rm d}t\,\exp[i(\omega_{ki}-\omega)t]/T$. Generally, when the pulse duration exceeds $n_p=10$ optical cycles, the Fourier coefficients converge rapidly to $C_{\omega}\rightarrow\delta(\omega-\omega_{\rm X})$ (cw limit \cite{watzel2019time}). Consequently, only continuum states with a fixed $E_k=\hbar\omega_{\rm X}+E_i$ are occupied for long pulses. Vice versa, for short pulses ($n_p<10$), we find that a distribution of continuum states with different kinetic energies contribute to the photoelectron wave packet.\\
Considering the matrix element $\langle\varphi^{(-)}_{\pmb{k}}|\pmb{A}_\omega(\pmb{r}_0)\cdot\pmb{r}|\Psi_i\rangle$, we note that the transverse and longitudinal vector potential components mediate different polarization states and the "ratio" between both depends on the atom's axial distance $\rho_0$ to the optical axis. Therefore, it is useful to separate the perturbative term into
\begin{equation}
\begin{split}
\pmb{A}_\omega(\pmb{r}_0)\cdot\pmb{r} &= \pmb{A}^{(\perp)}_\omega(\pmb{r}_0)\cdot\pmb{r}_\perp + A^{(z)}_\omega(\pmb{r}_0)z \\
&= \hat{Q}_\perp(\rho_0) + \hat{Q}_z(\pmb{r}_0)
\end{split}
\label{eq:LM_Hamiltonian}
\end{equation}
with
\begin{equation}
\hat{Q}_\perp(\pmb{r}_0) = -\sqrt{\frac{8\pi}{3}}\frac{\rho_0}{w_0}e^{-\rho_0^2/w_0^2}e^{-i\varphi_0}
\,rY_{1,1}(\Omega_{\pmb{r}}),
\label{eq:Qperp}
\end{equation}
and
\begin{equation}
\hat{Q}_z(\rho_0) = i\sqrt{\frac{4\pi}{3}}\frac{2}{kw_0}\left(1-\frac{\rho^2}{w_0^2}\right)e^{-\rho_0^2/w_0^2}
\,rY_{1,0}(\Omega_{\pmb{r}}).
\label{eq:Qz}
\end{equation}
Different selection rules for the angular quantum numbers apply   to transitions associated  with the transversal and longitudinal  contributions. For the former $\Delta L^{(\perp)}=1$ and $\Delta M^{(\perp)}=1$ applies; and for the latter $\Delta L^{(z)}=1$ and $\Delta M^{(z)}=0$.\\
The \emph{partial} photoionization amplitude corresponding to the transverse field component reads explicitly \cite{dahlstrom2013theory, kheifets2013time}
\begin{equation}
\begin{split}
\mathcal{A}^{(\perp)}_{i\pmb{k}}(\pmb{r}_0) =& (-1)^{m_i+1}\omega_{ki}C_{\omega_{ki}}F_{\perp}(\pmb{r}_0) \sum_{\ell=\ell_i\pm1} (-i)^\ell e^{i\delta_\ell(k)} \\
&\times \begin{pmatrix} \ell & 1 & \ell_i \\ -(m_i+1) & 1 & m_i \end{pmatrix} d_{\ell,n_i\ell_i}(k)Y_{\ell,m_i+1}(\Omega_{\pmb{k}}),
\end{split}
\end{equation}
where the reduced radial matrix elements are given by \cite{amusia2013atomic}
\begin{equation}
\begin{split}
d_{\ell,n_i\ell_i}(k)=&\sqrt{(2\ell+1)(2\ell_i+1)}
\begin{pmatrix} \ell & 1 & \ell_i \\ 0 & 0 & 0 \end{pmatrix} \\
&\times\int{\rm d}r\,rR_{k\ell}(r)R_{n_i,\ell_i}(r).
\end{split}
\end{equation}
The spatial distribution function is $F_{\perp}(\pmb{r}_0)=(\sqrt{2}\rho_0/w_0)\exp(-\rho_0^2/w_0^2-i\varphi_0)$.  The \emph{partial} photoionization amplitude associated with the longitudinal field component  reads
\begin{equation}
\begin{split}
\mathcal{A}^{(z)}_{i\pmb{k}}(\rho_0) =& i(-1)^{m_i}\omega_{ki}C_{\omega_{ki}}F_{z}(\rho_0) \sum_{\ell=\ell_i\pm1} (-i)^\ell e^{i\delta_\ell(k)} \\
&\times \begin{pmatrix} \ell & 1 & \ell_i \\ -m_i & 0 & m_i \end{pmatrix} d_{\ell,n_i\ell_i}(k)Y_{\ell,m_i}(\Omega_{\pmb{k}}),
\end{split}
\end{equation}
with the spatial distribution function $F_z(\rho_0)=(2/(kw_0))(1-\rho^2/w_0^2)\exp(-\rho_0^2/w_0^2)$.\\
The local (meaning $\pmb{r}_0$-dependent) time delay in photoionization  when  elevating the electron from its initial state $|i\rangle$ to the continuum is\cite{watzel2014angular}
\begin{equation}
\begin{split}
\tau^{n_i\ell_im_i}_{\rm at}(E_k,\Omega_{\pmb{k}},\pmb{r}_0)&=\frac{d}{dE_k}{\rm arg}\left[\mathcal{A}_{i\pmb{k}}(\pmb{r}_0)\right] \\
&=\Im\left[\frac{1}{k}\frac{\partial_k\mathcal{A}_{i\pmb{k}}(\pmb{r}_0)}
{\mathcal{A}_{i\pmb{k}}(\pmb{r}_0)}\right].
\end{split}
\end{equation}
The probability to observe the photoelectron follows from the partial differential cross section (DCS)
\begin{equation}
w^{n_i\ell_im_i}(E_k,\Omega_{\pmb{k}},\pmb{r}_0)=\left|\mathcal{A}_{i\pmb{k}}(\pmb{r}_0)\right|^2.
\label{eq:DCS}
\end{equation}
Its angular dependence reflects which angular channels are contributing and which interaction operator is dominating. Generally, the time delay and DCS are angular dependent when more (than one) angular channels are contributing to the total amplitude \cite{watzel2014angular} $\mathcal{A}_{i\pmb{k}}(\pmb{r}_0)=\mathcal{A}^{(\perp)}_{i\pmb{k}}(\pmb{r}_0) + \mathcal{A}^{(z)}_{i\pmb{k}}(\rho_0)$. However, as the driving laser field is three-dimensional, for an individual atom residing at $\pmb{r}_0$ the angular dependence is also determined by the presence (and ratio) of $\hat{Q}_\perp$ and $\hat{Q}_z$. If either $\hat{Q}_\perp$ or $\hat{Q}_z$ vanishes, $w^{n_i\ell_im_i}$ and $\tau^{n_i\ell_im_i}_{\rm at}$ exhibit cylindrical symmetry (referring to the atomic coordinate frame). If both are present, the ionization probability and time delay depend explicitly on $\vartheta_{\pmb{k}}$ and $\varphi_{\pmb{k}}$. \\
The local time delay associated with the whole subshell $n_i\ell_i$ follows from averaging over the magnetic sublevels {and} the bandwidth of the incident laser field, which is incorporated in the Fourier coefficients $C_{\omega_{ki}}$ in Eq.\,\eqref{eq:FourierC}, yielding
\begin{widetext}
\begin{equation}
 \tau_{\rm at}^{n_i\ell_i}(\Omega_{\pmb{k}},\pmb{r}_0)= \frac{\int{\rm d}E_k\,\sum_{m_i=-\ell_i}^{\ell_i} w^{n_i\ell_im_i}(E_k,\Omega_{\pmb{k}},\pmb{r}_0) \tau_{\rm at}^{n_i\ell_im_i}(E_k,\Omega_{\pmb{k}},\pmb{r}_0) } {\int{\rm d}E_k\,\sum_{m_i=-\ell_i}^{\ell_i} w^{n_i\ell_im_i}(E_k,\Omega_{\pmb{k}},\pmb{r}_0)}.
\end{equation}
\label{eq:td_avr}
\end{widetext}
The important quantity for an experimental realization is the relative time delay in photoionization  between two subshells, given by
\begin{equation}
\tau_{\rm at}^{n_a\ell_a-n_b\ell_b}(\Omega_{\pmb{k}},\pmb{r}_0)=\tau_{\rm at}^{n_a\ell_a}(\Omega_{\pmb{k}},\pmb{r}_0) - \tau_{\rm at}^{n_b\ell_b}(\Omega_{\pmb{k}},\pmb{r}_0).
\end{equation}
While the individual atom's $\tau_{\rm at}^{n_i\ell_i}(\Omega_{\pmb{k}},\rho_0)$ and DCS depend on $\pmb{r}_0$, in practice the sample is a thermal distribution of (non-interacting) atoms over the laser spot.  The thermal atoms are basically frozen on the time scale of the photoionization process so that the measured time-delay in photoionization amounts to the (incoherent) sample averaging. The average over the azimuthal coordinate
\begin{widetext}
\begin{equation}
 \overline{\tau}_{\rm at}^{n_i\ell_i}(\Omega_{\pmb{k}},\rho_0)= \frac{\int_0^{2\pi} {\rm d}\varphi_0 \int{\rm d}E_k\,\sum_{m_i=-\ell_i}^{\ell_i} w^{n_i\ell_im_i}(E_k,\Omega_{\pmb{k}},\rho_0) \tau_{\rm at}^{n_i\ell_im_i}(E_k,\Omega_{\pmb{k}},\rho_0) } {\int_0^{2\pi}{\rm d}\varphi_0\int{\rm d}E_k\,\sum_{m_i=-\ell_i}^{\ell_i} w^{n_i\ell_im_i}(E_k,\Omega_{\pmb{k}},\rho_0)}.
\label{eq:response_TD}
\end{equation}
\end{widetext}
helps  characterizing the dependency of the photoionization process on the axial radial distance $\rho_0$ of the atom to the optical axis.
We note that the integration over the azimuthal component of the atomic position vector $\pmb{r}_0$ lifts the azimuthal dependence of the ionization process (i.e., the dependency on $\varphi_{\pmb{k}}$) at the detector. Consequently, $\overline{\tau}_{\rm at}^{n_i\ell_i}(\Omega_{\pmb{k}},\rho_0)\equiv\overline{\tau}_{\rm at}^{n_i\ell_i}(\vartheta_{\pmb{k}},\rho_0)$. The full sample average is given by
\begin{widetext}
\begin{equation}
 \overline{\overline{\tau}}_{\rm at}^{n_i\ell_i}(\vartheta_{\pmb{k}})= \frac{\int_0^{2\pi}{\rm d}\varphi_0\int_0^\infty {\rm d}\rho_0\,\rho_0\int{\rm d}E_k\,\sum_{m_i=-\ell_i}^{\ell_i} w^{n_i\ell_im_i}(E_k,\Omega_{\pmb{k}},\rho_0) \tau_{\rm at}^{n_i\ell_im_i}(E_k,\Omega_{\pmb{k}},\rho_0) } {\int_0^{2\pi}{\rm d}\varphi_0\int_0^\infty {\rm d}\rho_0\,\rho_0 \int{\rm d}E_k\,\sum_{m_i=-\ell_i}^{\ell_i} w^{n_i\ell_im_i}(E_k,\Omega_{\pmb{k}},\rho_0)}.
\label{eq:response_TD_full}
\end{equation}
\end{widetext}

\section{Polarization-dependent atomic time delay in argon}
\label{sec:at_td}
Time delay in photoionization is well studied for argon atoms. $\tau_{\rm at}^{3p-3s}$ exhibits strong angular variation for photon energy  $\approx$ 50\,eV, i.e., around the Cooper minimum for the 3p photoionization cross section  \cite{watzel2014angular, dahlstrom2014study}, in which case  the relative magnitudes of the $3p\rightarrow ks$ and $3p\rightarrow kd$ are comparable   resulting in  angular variation of the quantum phase. Here, we consider Ar atoms  within an effective  single-particle model potential \cite{sarsa2004parameterized} that  reproduces the 3p Cooper minimum with reasonable  accuracy (to reproduce the 3s Cooper minimum for lower XUV energies   accounting  for many-body effects \cite{kheifets2013time,dixit2013time} is necessary).
The  time delay is  sensitive to the distance $\rho_0$  as long as several angular channels are participating. The reason for this behavior  is that the ratio between the contributing angular channels varies with  the relative strengths of $\hat{Q}_\perp$ and $\hat{Q}_z$.
\begin{figure}[t]
\centering
\includegraphics[width=8.4cm]{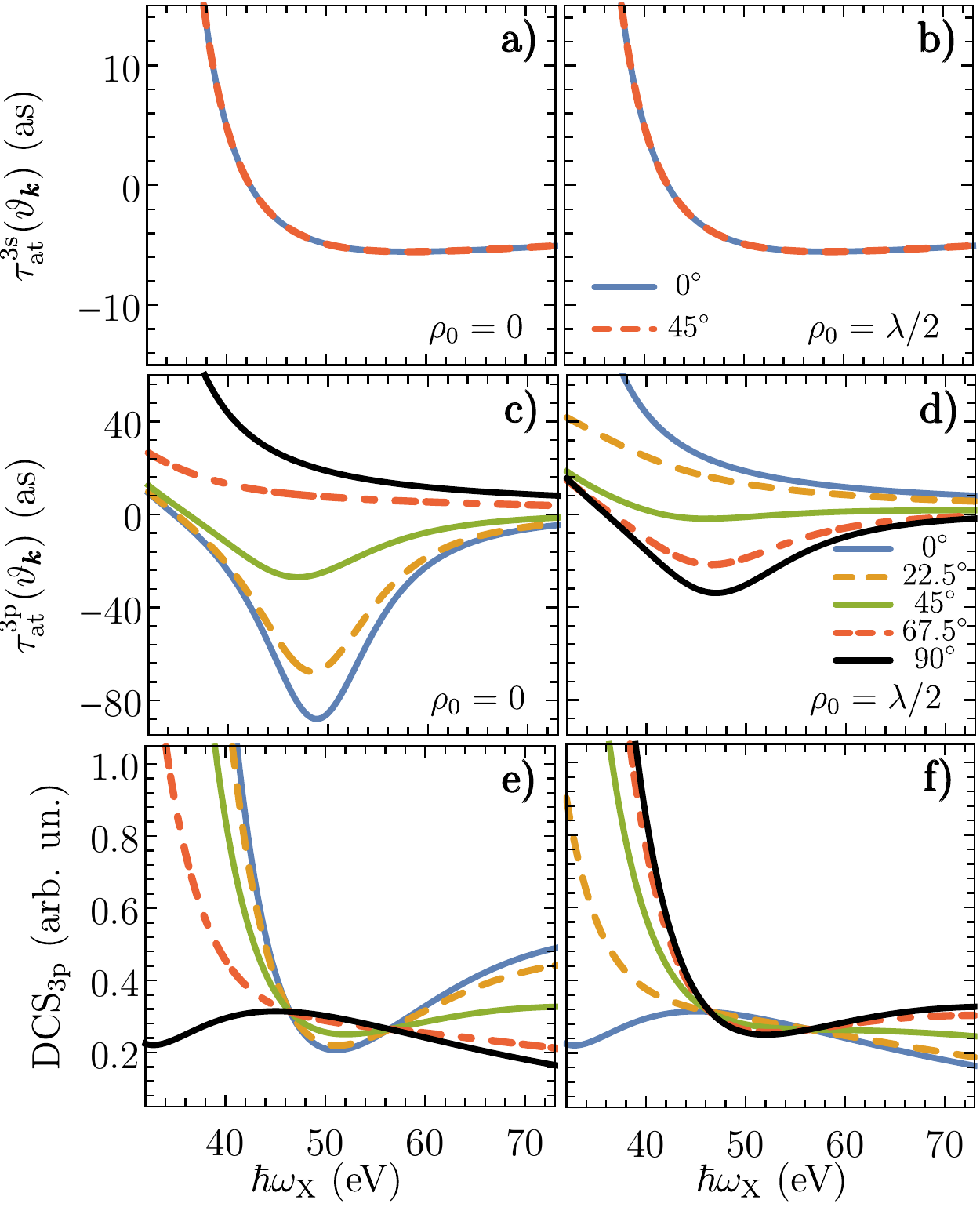}
\caption{Energy and time-dependent subshell time delays $\tau_{\rm at}^{\rm 3s}$ (upper row) and $\tau_{\rm at}^{\rm 3p}$ (lower row) for different atom axial distances $\rho_0$ away from the optical axis. The last row shows the corresponding ionization probabilities following Eq.\,\eqref{eq:DCS}.}
\label{Fig2}
\end{figure}
 Fig.\,\ref{Fig2} shows the subshell time delays $\tau_{\rm at}^{\rm 3s}$ and $\tau_{\rm at}^{\rm 3p}$ for Ar atom located at the optical axis (left column) and at $\rho_0=\lambda/2$ (right column).
\begin{figure}[b]
\centering
\includegraphics[width=8.4cm]{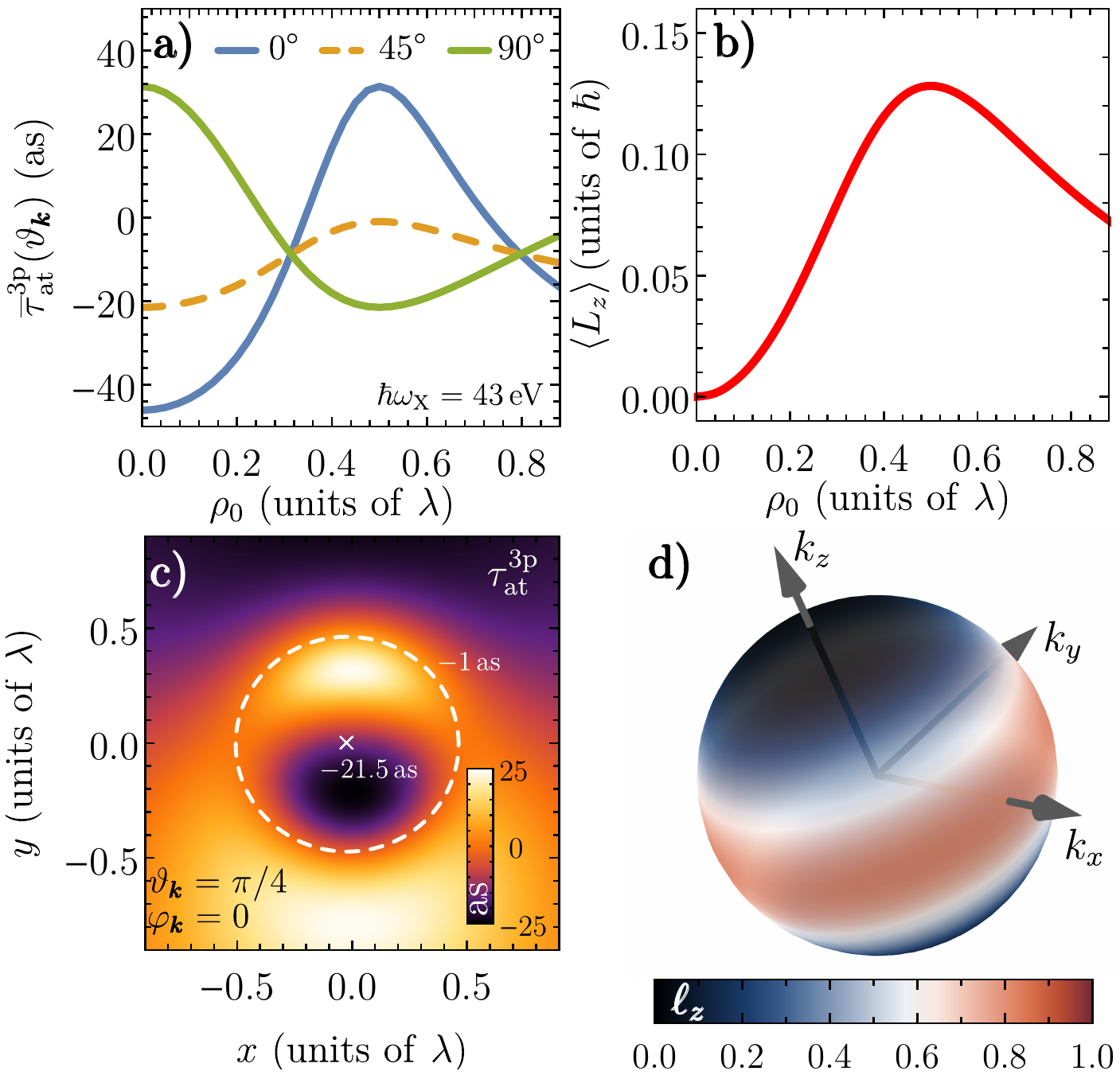}
\caption{a) Atomic time delay as a function of the axial distance $\rho_0$ for different observation angles $\vartheta_{\pmb{k}}$ and a fixed photon energy of the incident laser field. b) Acquired angular momentum $\langle L_z\rangle$ of the photoelectron. b) Full spatial dependence of the 3p time delay in the laser spot for $\vartheta_{\pmb{k}}=45^\circ$. The white dashed circle indicates the averaged time delay for a radius of $\rho_0=0.5\lambda$ (-1\,as). d) Angular momentum density $\ell_z$ in dependence on the asymptotic momentum $\pmb{k}$ for a photoelectron launched from $\rho_0=\lambda/2$.}
\label{Fig3}
\end{figure}
For $\rho_0=0$ ($\hat{Q}_\perp=0$) the dominating longitudinal  component (of the radial beam) acts locally as a linear polarized field, while for $\rho_0=\lambda/2$ the transversal field component of $\pmb{A}(\pmb{r},t)$ is dominating ($\hat{Q}_z=0$) and exhibits circular polarization. Since only one angular channel ($k$p) is available for the photoelectron, the 3s time delay shows no angular dependence (regardless of $\rho_0$).  In contrast, the 3p time delay exhibits  a manifold dependence on both the emission angle $\vartheta_{\pmb{k}}$ and the atomic position (distance $\rho_0$), which manipulates the interference between the $k$s and $k$d angular channels.\\
In Fig.\,\ref{Fig2}c), the depicted curves follow the well-known trend of the atomic time delay in the vicinity of the 3p Cooper minimum, where a strong angular variation can be found \cite{watzel2014angular}. At the propagation direction of the incident optical field ($\vartheta_{\pmb{k}}=0^\circ$), the time delay is strongly negative \cite{kheifets2013time}, while observing the photoelectron perpendicular (to the axis) yields a positive $\tau_{\rm at}^{\rm 3p}$. The situation changes markedly  when switching the polarization state to circular (by investigating an Ar atom located at $\rho_0=\lambda/2$), which is presented in Fig.\,\ref{Fig2}d). The subshell time delay $\tau_{\rm at}^{\rm 3p}$ is now strongly negative when observing the photoelectron perpendicular to the propagation direction and turns positive in direction $\vartheta_{\pmb{k}}=0^\circ$. Note that the time delay in this field region can be well described by the theory presented in Ref.\,\cite{ivanov2013time}. In both cases, the time delay exhibits cylindrical symmetry, i.e., it is invariant to variation in the (observation) azimuthal angle $\varphi_{\pmb{k}}$ or to the exact position on the circle with radius $\rho_0=\lambda/2$. \\
As inferred from Fig.\,\ref{Fig1}b), the incident vector beam enables a smooth transition between the limiting cases shown in Fig.\,\ref{Fig2}, when scanning the time delay over the axial distance $\rho_0$.  For intermediate positions (between $\rho_0=0$ and $\rho_0=\lambda/2$) when both $\hat{Q}_\perp$ and $\hat{Q}_z$ are present, we have to account for variations on $\varphi_{\pmb{k}}$ and $\varphi_0$. If one is interested in the dependence on the axial distance $\rho_0$, it is appropriate to consider the \emph{averaged} time delay of atoms located on a circle with the radius $\rho_0$, i.e., using Eq.\,\eqref{eq:response_TD} for a fixed axial distance. The numerical integration reveals that the averaged (but $\rho_0$-resolved) time delay is cylindrically symmetric, i.e., it depends only on the polar angle $\vartheta_{\pmb{k}}$. In Fig.\,\ref{Fig3}a), the evolution of the \emph{circle-averaged} 3p subshell time delay versus $\rho_0$ is presented for different emission angles $\vartheta_{\pmb{k}}$. In all cases, $\tau_{\rm at}^{\rm 3p}$ changes its sign when entering the area where the transversal field component of $\pmb{A}(\pmb{r},t)$ starts to dominate (i.e., $\rho_0>0.3\lambda)$. Therefore, in practically all emission angles, we find a strong sensitivity of the time delay to the space-dependent field and polarization state distribution.
From the photoionization of the valence shell electron with the structured light field one infers  a dependence of the time delay on the azimuthal coordinate $\varphi_0$ in the laser spot. It is pronounced for polar scattering angles $\vartheta_{\pmb{k}}$, where the interplay between $\hat{Q}_\perp$ and $\hat{Q}_z$ is sustainable (e.g. $\vartheta_{\pmb{k}}=\pi/4$). Such a scenario is depicted in Fig.\,\ref{Fig3}c) which  shows the 3p time delay in the whole beam spot. We observe a periodicity that reflects the phase $\exp(-i\varphi_0)$ in $\hat{Q}_\perp$. Furthermore,  the entire structure can be rotated by changing the azimuthal observation angle $\varphi_{\pmb{k}}$. The corresponding circle-averaged results are presented in Fig.\,\ref{Fig3}a), as demonstrated, for instance, by the dashed circle with radius $\rho_0=0.5\lambda$ which belongs to an averaged time delay $\overline{\tau}_{\rm at.}^{3p}=-1$\,as. Therefore, the proposed structured laser scheme also allows identifying the atomic position  with respect to the azimuthal coordinate $\varphi_0$ in the laser spot. However, we note that substantial variations of the time delay (i.e., with a change of sign) with $\phi_0$ only occur for axial distances $\rho_0$ where $\hat{Q}_\perp$ and $\hat{Q}_z$ are in the same magnitude, e.g., at $\rho_0=0.3\lambda$ or $\rho_0>0.7\lambda$  (cf.\,Fig.\,\ref{Fig1}). Typically, the total ionization probability in these cases is reduced.\\
Since the field polarization changes locally within the laser spot from linear to circular, it is of relevance to look into the acquired angular momentum (shown in Fig.\,\ref{Fig3}b), which is then expected to be dependent on the atomic position. It can be found by the computation of the expectation value of the $z$ component of the angular momentum operator $\hat{L}_z=-i\partial_\varphi$:
\begin{equation}
\ell_z(\pmb{k},\rho_0) = -i\int{\rm d}\varphi_0\sum_{m_i=-\ell_i}^{\ell_i}\mathcal{A}_{\pmb{k}i}^*(\pmb{r}_0)\,\partial_\varphi \mathcal{A}_{\pmb{k}i}(\pmb{r}_0),
\end{equation}
while $\langle L_z\rangle(\rho_0) = \int{\rm d}\pmb{k}\,\ell_z(\pmb{k},\rho_0)$. Near the optical axis, $\langle L_z\rangle$ vanishes  (dominant longitudinal field  component is  linearly polarized),  increasing with  $\rho_0$.  Recalling Fig.\ref{Fig1}, the laser field is circularly polarized in a narrow region (well below the diffraction limit).
Thus, if we were to access $\langle L_z\rangle$ experimentally, we would have  a spatial resolution on the initial state of the photoelectrons below the optical diffraction limit. In this context we note: i), no net orbital angular momentum is carried by the radial or by the  azimuthal vector beam \cite{watzel2019magnetoelectric}. The finite $\langle L_z\rangle$ is due to the coherent superposition of both. ii), a finite $\langle L_z\rangle$ implies a finite linear momentum component in he transversal direction which can be measured. Another indication of finite $\langle L_z\rangle$ is
the correlation with the time delay $\tau_{\rm at}^{\rm 3p}$ and its   sign change. At the peak of $\langle L_z\rangle$ (where the impact of $\pmb{A}_{\perp}(\pmb{r},t)$ is maximal), the atomic time delay has an  extremum. \\
To compare with experiments averaging over the atom positions $\pmb{r}_0$ is necessary. Fig.\,\ref{Fig4} shows an analysis of the contributions to the sample-averaged  3p-3s time delay $\overline{\tau}^{\rm 3p-3s}_{\rm at}$: The black curve corresponds  to the case where only the longitudinal  contribution $\hat{Q}_z$ is considered, meaning that  only $z$-linear polarization is present. The whole response of the gas is then similar to photoionization with a homogeneous laser field and the time delay is identical to the well-known results (see, for instance, Refs.\,\cite{watzel2014angular}). The grey, dashed curves represent the other extreme case, where only the transversal contribution $\hat{Q}_\perp$ is present. Experimentally, this situation can be realized by a weak focussing condition, suppressing  effectively the longitudinal component. The action of $\hat{Q}_\perp$ is reminiscent to  photoionization with circularly polarized fields, and  the properties of the time delay can be explained by the theoretical model developed in Ref.\,\cite{ivanov2013time}. Note  the opposite signs of the relative time delays $\overline{\tau}^{\rm 3p-3s}_{\rm at}$ are associated with either $\hat{Q}_\perp$ or $\hat{Q}_z$.\\
The other curves show the sample-averaged time delay for different sizes of the (disc-type) atom distributions when the full field is acting. The distribution with the smallest radius of $\lambda/8$ (orange, dashed curve) is practically identical to the result for purely linear polarized light (black curve). This observation is consistent with Fig.\ref{Fig1}, leading to the conclusion that if such a distribution of the time-delay is observed, then the atom resides near the optical axis. In principle, a case of very dilute cold atom target was realized experimentally \cite{schmiegelow2016transfer}.
\begin{figure}[t]
\centering
\includegraphics[width=8.4cm]{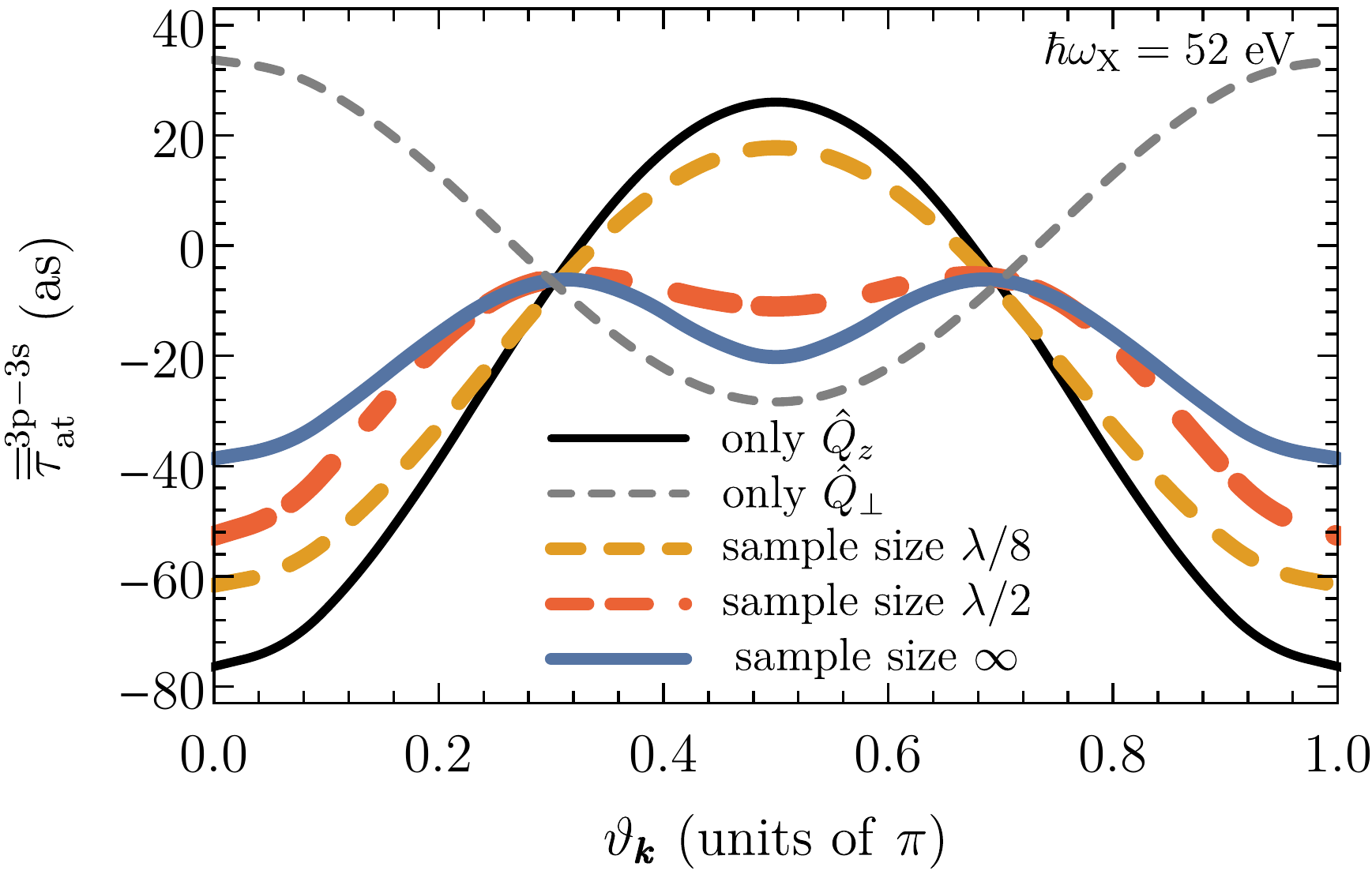}
\caption{Relative atomic time delay, statistically averaged over the atoms distribution, as a function  of the observation angle $\vartheta_{\pmb{k}}$. Black and grey, dashed curve show respectively the results when only $\hat{Q}_z$ or $\hat{Q}_\perp$  are present in the light-matter interaction Hamiltonian, Eq.\,\eqref{eq:LM_Hamiltonian}. The colored curves correspond  to different sample sizes of the disk-shaped atomic distributions.}
\label{Fig4}
\end{figure}
Enlarging the gas area (red dot-dashed curve) increases the influence of the transversal contributions with circular polarization. In the propagation direction ($\vartheta_{\pmb{k}}=0$), the relative delay increases while it shrinks in perpendicular observation direction ($\vartheta_{\pmb{k}}=\pi/2$). For infinitely large gas sample (few times larger than the  laser spot, shown by the blue curve), we observe that in contrast to the cases where either $\hat{Q}_\perp$ (black curve) or $\hat{Q}_z$ (gray curve) are dominating the interaction, the $\overline{\tau}^{\rm 3p-3s}_{\rm at}$ has no sign change, but minima in propagation ($\vartheta_{\pmb{k}}=0$) and transverse direction ($\vartheta_{\pmb{k}}=\pi/2$) emerge. comparing with the black and gray curves, we identify the minimum at $\vartheta_{\pmb{k}}=0$ originating from atoms located at and around the optical axis. On the other hand, the minimum at $\vartheta_{\pmb{k}}=\pi/2$ is the effect of photoelectrons launched from positions in the laser spot with circular polarization, i.e., from atoms located at and around a ring with radius $\rho_0=\lambda/2$.
\section{Spatial identification via structured light based on attosecond streaking}

\subsection{photoionization probability}

\begin{figure}[t!]
\includegraphics[width=8.4cm]{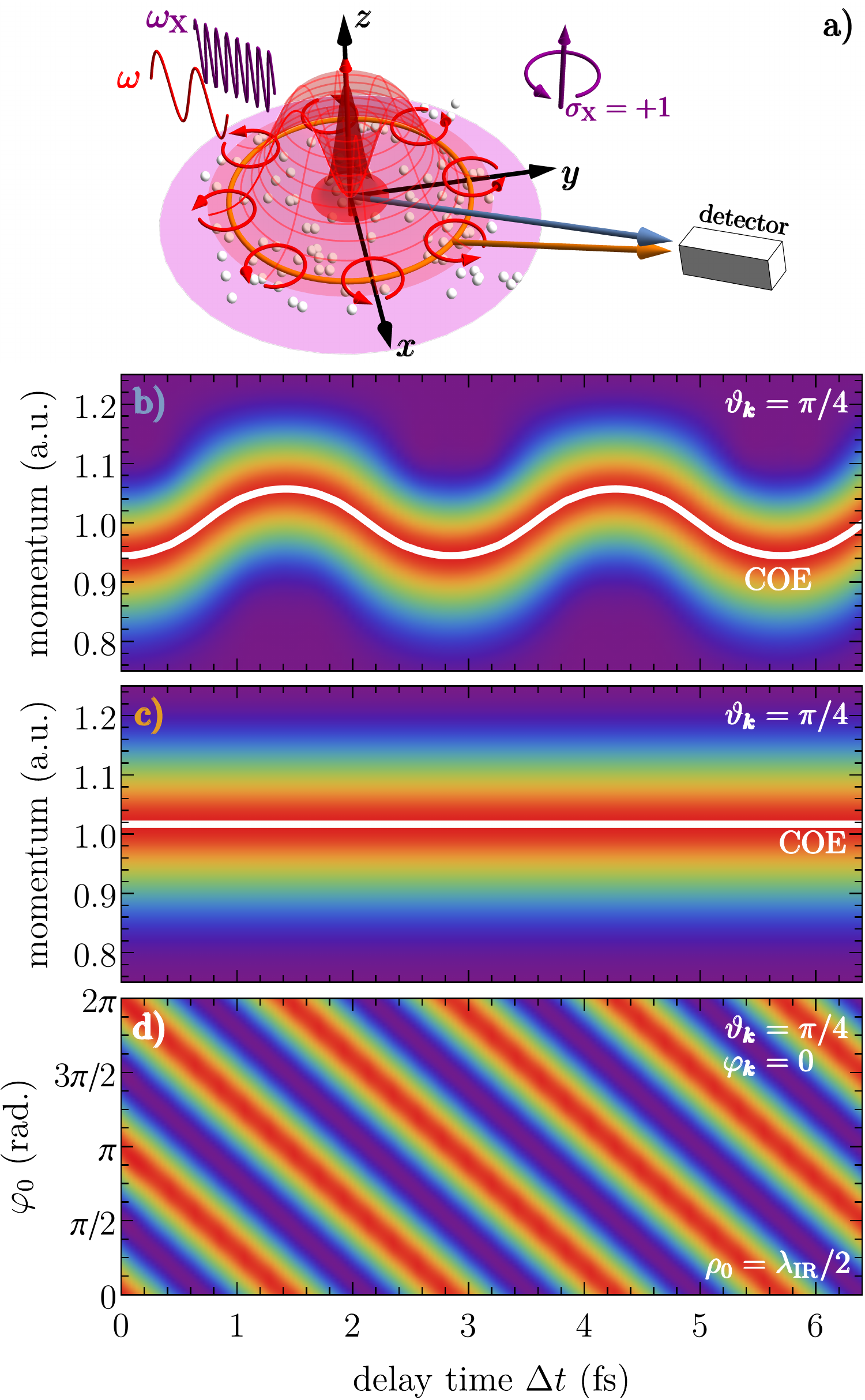}
\caption{a) Schematic representation of the structured IR streaking laser applied to atoms in addition to an ionizing, unstructured, circularly polarized XUV laser. Depending on the spatial origin of the atom, we observe different streaking behavior. b) Streaking spectrum corresponding to an atom located at the optical axis. c) (Averaged) streaking spectrum of atoms located on a circle in the vicinity of the intensity maximum at $\rho_{0}=\lambda_{\rm IR}/2$. d) Color map of the streaking signal for a fixed final kinetic momentum $k=1$\,a.u. as a function of the time delay $\Delta t$ and the azimuthal coordinate $\varphi_0$ of an atom residing on a circle with the radius $\lambda_{\rm IR}/2$.}
\label{FigStr1}
\end{figure}

To access phase information on the photoelectron, one applies a second laser field with the vector potential $\pmb{A}_{\rm L}(\pmb{r},t)$ and the energy $\hbar\omega_{\rm L}$. The contributing amplitudes to the final kinetic energy state $|\varphi_{\pmb{k}}^{(-)}\rangle$ of the photoelectron have different quantum phases. The detected photoionization probability is therefore influenced by interference which shows up in time-dependent cross terms. The temporal information on the photoionization in the presence of both laser fields can then  be mapped onto the energy scale and the time delay can be extracted. The most common techniques, attosecond streaking \cite{goulielmakis2004direct} and RABBIT \cite{klunder2011probing}, rely on this physical mechanism and provide similar time information.\\
In the context of our studies, attosecond streaking serves not only as a tool for extracting phase information on the photoelectrons, but itself can be used for spatial identification on the sub-wavelength scale. In the following, we consider collinearly propagating XUV und IR laser fields. The pump pulse is unstructured and circularly polarized with the helicity  $\sigma_{\rm X}=+1$. The intensity is $5\times10^{13}$\,W/cm$^2$ and the pulse length $T_{\rm X}$ accommodates eight optical cycles. The photon energy in our example is $\hbar\omega_{\rm X}=30$\,eV. The corresponding Keldysh parameter is $\gamma>1$, which is typical for experiments with XUV laser fields \cite{kheifets2010delay, de2020photoelectric}. The XUV laser field ionizes a dilute gas of Ar atoms into the vector potential $\pmb{A}_{\rm L}(\pmb{r},t)$ of a structured IR field with the analytical properties given in Sec.\,\ref{sec:I}. The pulse length is 10 optical cycles, $\hbar\omega=1.58$\,eV, and the peak intensity at $w_0$ is $5\times10^{11}$\,W/cm$^2$, which is typical for streaking experiments \cite{schultze2010delay}. The waist of the structured IR field is $\omega_0=392$\,nm ($\lambda_{\rm IR}/2$). A schematic representation of the laser setup is shown in Fig.\,\ref{FigStr1}a).\\
Recently, we developed a variant of the strong field approximation (SFA) for structured light fields \cite{waetzel2020Electrons}, which we apply to obtain the streaking dynamics via numerical time integration. The transition amplitude for an atom located at $\pmb{r}_0$ is
\begin{equation}
\begin{split}
\mathcal{A}_{i\pmb{k}}(\pmb{r}_0,\Delta t)=&i\int_{-\infty}^{\infty} {\rm d}t\,\langle\varphi^{(-)}_{\pmb{k}}|e^{i\pmb{A}_{\rm L}(\pmb{r}_0,t)\cdot\pmb{r}}rY_{1,1}(\Omega_{\pmb{r}})|\Psi_i\rangle \\
&\times \mathcal{E}_{\rm X}(t-\Delta t)e^{i(S_V(\pmb{r}_0,\pmb{k},t) + I_pt)}.
\end{split}
\label{eq:SFA}
\end{equation}
$\mathcal{E}_{\rm X}(t)=f(t)\cos(\omega_{\rm X}t)$ is the temporal function of the XUV electric field with the frequency $\omega_{\rm X}$ and $f(t)=\sin(\pi t/T_{\rm X})^2$ for $t\in[0,T_{\rm X}]$. The function $S_V(\pmb{r}_0,\pmb{k},t)$ is the position-dependent modified Volkov phase for the structured light fields while $I_p$ is the ionization potential. For the explicit representation of the Volkov phase for vector beams (and other structured light fields),
we refer to Ref.\,\cite{waetzel2020Electrons}. The parameter $\Delta t$ is the temporal delay between the maxima of the XUV and streaking IR fields. Varying $\Delta t$ gives an insight into the temporal aspects of the photoionization dynamics.\\
Fig.\,\ref{FigStr1} shows results of two-color ionization within the formalism of Eq.\,\eqref{eq:SFA}, where the XUV-field is unstructured and circularly polarized, while the liberated electron is exposed to the collinearly propagating structured laser-assisting field described by $\pmb{A}_{\rm L}(\pmb{r},t)$. We study the \emph{circle-averaged} streaking spectra $\int_0^{2\pi}{\rm d}\varphi_0\,|\mathcal{A}_{i\pmb{k}}(\pmb{r}_0,\Delta t)|^2$ for an atom located at the optical axis ($\rho_0=0$) and at the radial intensity maximum at $\rho_{\rm max}$. The signal is averaged over the atomic positions on a circle with the radius $\rho_{\rm max}$. While the former is exposed to a strong longitudinal component, the streaking dynamics of the latter is determined by the transversal component of $\pmb{A}_{\rm L}$, which is circularly polarized. However, the interaction with matter depends on the azimuthal position due to the presence of the phase $\exp(i\varphi_0)$ (cf.\,Eq.\,\eqref{eq:Qperp}).\\
Positioning now the photoelectron detector at $\vartheta_{\pmb{k}}=\pi/4$ (relative to the propagation axis of both light fields), we may observe different streaking behavior depending on the (spatial) origin of the ionized atom as shown by the panels b) and c). The (streaked) electrons launched from atoms in the center of the beam spot exhibit the well-known streaking spectra with the delay-modulated ionization probability, as presented by Fig.\,\ref{FigStr1}b). Interestingly, the modulation disappears in the case of the (averaged) response of photoelectrons originating from atoms around the high-intensity rim where the transversal component of $\pmb{A}_{\rm L}(\pmb{r},t)$ dominates. This can be explained by the phase factor $\exp(i\varphi_0)$ in $\pmb{A}_{\rm L}(\pmb{r},t)$ (see Eq.\,\eqref{eq:Qperp}) which results in averaging out the modulation (which is present when considering only one atom residing on the circle with radius $\rho_{\rm max}$). The influence of $\varphi_0$ (and the resulting averaging) is highlighted by the color map in Fig.\,\ref{FigStr1}d), where the (two-color) ionization probability is shown for a fixed kinetic momentum in the dependence on $\Delta t$ and $\varphi_0$.

\subsection{Streaking time delay}
\begin{figure}[t!]
\includegraphics[width=8.4cm]{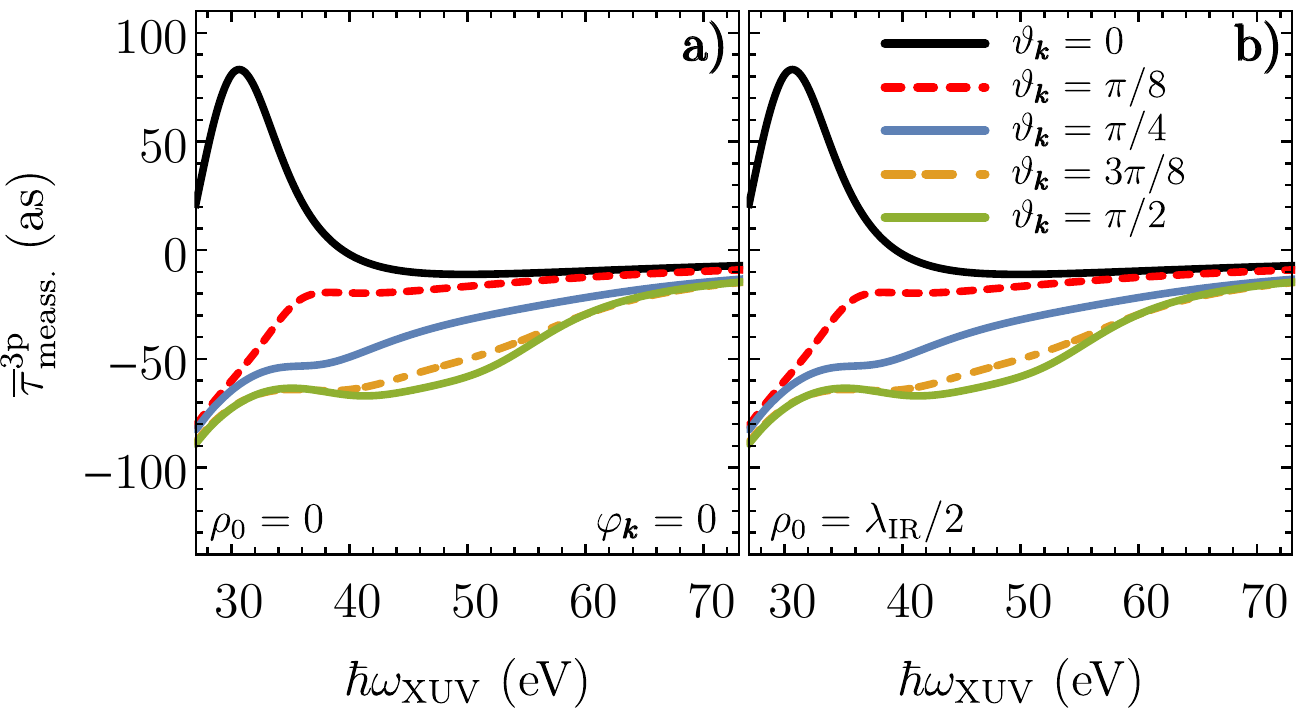}
\caption{Time delay for Ar $3p$ subshell as a function of  the photon energy of the XUV field and the asymptotic emission angle $\vartheta_{\pmb{k}}$. Left column: time delay for an atom located at the optical axis of the structured streaking field $\pmb{A}_{\rm L}(\pmb{r},t)$. Right column: averaged time delay for atoms located on a circle with a radius $\rho_0=\lambda_{\rm IR}/2$ (at the donut rim). The laser parameters are the same as described in main text.}
\label{FigStr2}
\end{figure}

What is the impact of structuring the streaking field on the time delay?. In general, the (measured) time delay $\tau_{\rm meass.}$ can be separated into an intrinsic part, the atomic time delay $\tau_{\rm at}$ as covered in section Sec.\,\ref{sec:at_td}, and an extrinsic contribution $\tau_{\rm cc}$ stemming from the interaction of the liberated electron with the assisting laser field L \cite{nagele2011time, dahlstrom2013theory, pazourek2013time}. To be more precise, while the $\tau_{\rm at}$ reflects the atom-specific scattering characteristics, $\tau_{\rm cc}$ depends mainly on the external (experimental) parameters such as the amplitude $A_0^{\rm L}$ (at the atomic position $\pmb{r}_0$), and the frequency $\omega_{\rm L}$.  The treatment of the continuum-continuum (cc) transitions, driven by $A_0^{\rm L}$, in the framework of perturbation theory shows  that $\tau_{\rm cc}$ is independent of the intermediate and final angular momenta states of the photoelectron \cite{dahlstrom2014study}. This implies an independence on the polarization state of the streaking field,
as evidenced  by Fig.\,\ref{FigStr2} showing  the time delays as extracted from the numerical propagation \cite{nurhuda1999numerical} of the 3D Schr\"{o}dinger equation in a parametrized single-particle potential \cite{sarsa2004parameterized} and the vector potentials of the unstructured XUV and structured laser-assisting field. To obtain $\tau_{\rm meass.}$ corresponding to an atom located at $\pmb{r}_0$, the COEs of the numerically obtained streaking spectra as functions of the delay time $\Delta t$ are fitted to $\pmb{k}_f(\Delta t)=\pmb{k}_0-\alpha\pmb{A}_{\rm L}(\pmb{r}_0,\Delta t + \tau_{\rm meass.})$, where $k_0=\sqrt{2(\omega_{\rm X}+E_i)}$. The left panel in Fig.\,\ref{FigStr2} corresponds to an Ar atom located at the optical axis ($\pmb{r}_0=0$), where the liberated $3p$-electron (by a $z$-linearly polarized XUV field) is primary streaked by the strong longitudinal component of the structured IR field. In contrast, the right panel shows a streaked photoelectron originating from an atom located in the intensity donut of the streaking field, which is characterized by a dominating transversal component (which is locally circularly polarized). The differences between the particular curves are marginal emphasizing the universality of $\tau_{\rm cc}$ with respect to the emission angle and polarization state. In other words, the cc-contribution to the measured time delay is robust to the spatial structuring of the streaking field. \\
This robustness has interesting implications for mixtures of atomic gases. While $\tau_{\rm at}$ is atom-specific and $\tau_{\rm cc}$ is universal, the ionization streaking probability depends crucially on the position of the atom within the laser spot, as presented in Fig.\,\ref{FigStr1}. Therefore, we can not only identify the origin of the measured photoelectron but also the atom type via the characteristic $\tau_{\rm at}$.
\section{Conclusions}
A superposition of focused radial and azimuthal vector beams results in a photonic field that is longitudinal and linearly polarized at the beam center and transversally circularly polarized at the boundary of the laser spot. The relative ratio of these two components across the beam spot depends on the focusing of the radial beam and its intensity. Photoionizing an Ar atom with the combined laser field, we find that the photoelectron  angular distribution, the quantum phase, and the time delay of a liberated valence shell electron depend on the atom position. For instance, Photoelectrons launched at the beam center and observed in field propagation direction exhibit a negative time delay. If, however, the ionized atom is at the laser spot rim, the time delay is positive. Similar spatially dependent signatures are observed when instead of the (ionizing) XUV laser, the streaking field is polarization-structured. Interestingly, the modulation of the streaking spectrum disappears when looking at the averaged response of atoms located in the intensity rim of the structured streaking field.

\begin{acknowledgements}
This study is supported by the Deutsche Forschungsgemeinschaft
(DFG) under SPP1840, and WA 4352/2-1. We thank the anonymous reviewers for valuable suggestions.
\end{acknowledgements}

%

\end{document}